\documentclass[12pt]{article}
\usepackage{graphicx}
\usepackage{amssymb}
\usepackage{amsmath}
\usepackage{epstopdf}
\usepackage{hyperref}

\newcommand{\di}{\mathrm{i}}

\newcommand{\D}{{\mathcal D}}
\title{
Octonions, Monopoles, and Knots}
\author{
Sergey A. Cherkis\\
\\
\it Department of Mathematics\\
\it University of Arizona\\
\it Tucson AZ, 85721-0089, USA\\
\tt cherkis@math.arizona.edu}
\date{}
                                         
\begin{document}
\begin{titlepage}

\renewcommand{\thepage}{ }
\maketitle
\abstract{Witten's approach to Khovanov homology of knots is based on the five-dimensional system of partial differential equations, which we call Haydys-Witten equations.  We argue for a one-to-one correspondence between its solutions  and solutions of the  seven-dimensional system of equations.  The latter can be formulated on any G2 holonomy manifold and is a close cousin of the monopole equation of Bogomolny.  Octonions play the central role in our view, in which both the seven-dimensional equations  and the Haydys-Witten equations appear as reductions of the eight-dimensional Spin(7) instanton equation.} 

\end{titlepage}


\section{Introduction}
In this note we propose a dual description of the Haydys-Witten equations \eqref{Eq:HW}, that originally appeared in \cite[Eq.(14)]{Haydys:2010dv}  and \cite[Eq.(5.36)]{Witten:2011zz}. They play a central role in Witten's  categorification of the Jones polynomials \cite{Witten:2011zz}, which is equivalent \cite{Witten:2011pz} to that of  Khovanov \cite{Khovanov99,Khovanov01}.  These equations are related to an earlier work of Pidstrigach \cite{P} generalizing the Seiberg-Witten equations and to the gauge-theoretic description of the string theory configuration of \cite{Gukov:2004hz} and \cite{Ooguri:1999bv}.

The Haydys-Witten equations are formulated in five dimensions.  They are particularly interesting, as they incorporate a number of other important systems of equations such as the Kapustin-Witten equations \cite{Kapustin:2006pk}, the Vafa-Witten equations \cite{Vafa:1994tf}, and the self-dual Yang-Mills equation with all of its reductions (such as, the Bogomolny equation, the Hitchin System, and the Nahm equation).  Here we demonstrate that the Haydys-Witten equations  in turn can be viewed as a result of  the reduction of the Spin(7) instanton equation of \cite{CDFN} and \cite{Ward}, more recently studied in \cite{ReyesCarrion:1998si} and \cite{Donaldson:1996kp, Donaldson:2009yq}.

We review the Haydys-Witten system in Section~\ref{HaydysWitten}.  This is a system of seven equations in five dimensions that we write concisely using octonions in Section~\ref{Sec:Octonions}.  This formulation makes it clear that it descends from an eight-dimensional system of equations known as a Spin(7) instanton equations to five dimensions via a straightforward dimensional reduction.  One can explore various other reductions of Spin(7) instanton equations; among these it is the reductions to seven and to three dimensions that are most relevant for our purposes.  Both are expected to be dual to the five-dimensional system and both appear in Section~\ref{Sec:Octonions}.  Given the flat space equations and their interpretation as Spin(7) instanton reduction, we seek their general covariant form.  For the seven-dimensional equation to make sense the underlying space has to possess a G2 structure.  After a brief review of various facts related to special holonomy in Section~\ref{Sec:Covariant}, we formulate in Section 5 the seven-dimensional equation on any manifold with G2 holonomy:
\begin{equation}
*\left(\psi^{(4)}\wedge F_A\right)=-D\Phi.
\end{equation}
This equation is rooted in octonions and has a close resemblance of the Bogomolny monopole equation in three dimensions.  Thus we call it the {\em octonionic monopole equation}.    

In the remainder of the text we explore the relation between the octonionic monopole in seven dimensions and the Haydys-Witten equation in five dimensions.    If the latter are studied on $\mathbb{R}\times W_3\times\mathbb{R}_+$, for example, where $W_3$ is a three-manifold, then the former should be considered on $\mathbb{R}\times T^*W_3,$ where $T^*W_3$ is the cotangent bundle to this three manifold. Section~\ref{Sec:Branes} gives a string theoretic reason to expect such a relation.  Moreover, the string theory picture allows one to identify the boundary conditions that correspond to the introduction of a knot $K\subset W_3.$  
The concluding section discusses the seven-dimensional interpretation of the knot invariants.

\section{Haydys-Witten Equations}\label{HaydysWitten}
The five-dimensional equations  of \cite{Haydys:2010dv}  and \cite{Witten:2011zz} involve a connection, with the gauge potential one-form $A,$ on a rank $n$ Hermitian vector bundle $E$ over a five-dimensional manifold $M_4\times\mathbb{R}_+$ (with local coordinates on $M_4$ being $x^\mu, \mu=0,1,2,3$ and the factor $\mathbb{R}_+$ parameterized by $y\geq0$) and a $y$-dependent self-dual two-form $B=B_{\mu\nu}dx^\mu dx^\nu$ on $M_4$ valued in the adjoint bundle ${\rm ad}(E)$:  
\begin{align}
A&\in\Omega^1(M_4\times\mathbb{R}_+)\otimes{\rm ad}(E)&& \text{and} &B&\in\Omega^{2,+}(M_4)\otimes{\rm ad}(E).
\end{align}

Of particular interest is the case of $M_4=\mathbb{R}\times W_3.$  If $x^\mu$ are local coordinates on $M_4,$ we choose $x^i,\, i=1,2,3$ to be the local coordinates on $W_3$ and $x^0$ or $t$ to be the coordinate on $\mathbb{R}.$
The self-dual two-form field $B=\frac{1}{2\sqrt{2}}B_{\mu\nu}dx^\mu\wedge dx^\nu$ has three independent components and for the flat case of $W_3=\mathbb{R}^3$ these can be written as
\begin{align}\label{Eq:B}
B_{0i}&=\phi_i, & B_{ij}&=\epsilon_{ijk}\phi_k.
\end{align}
Here $\epsilon_{ijk}$ is the totally antisymmetric tensor with $\epsilon_{123}=1.$ 
In general, if $\omega_1, \omega_2,$ and $\omega_3\in\Omega^{2,+}(M_4)$ form an orthonormalized basis of self-dual forms in each fiber of $\Omega^{2,+}(M_4)$, we can decompose the field $B$ as $B=\phi^k\omega_k$ and define $B\times B=(\vec{\phi}\times\vec{\phi})^k\omega_k=\epsilon_{ijk}\phi^i\phi^j\omega_k.$ 
The central system of equations of \cite{Haydys:2010dv} and \cite{Witten:2011zz} is the Haydys-Witten system
\begin{subequations}
\label{Eq:HW}
\begin{align}\label{Eq:Main}
F^+-\frac{1}{4} B\times B-\frac{1}{2} D_y B&=0,\\
\label{Eq:Main2}
F_{y\mu}+D^\nu B_{\nu\mu}&=0.
\end{align}
\end{subequations}
Here  $F=dA+A\wedge A$ is the  curvature of the connection, and $F^+$ is the self-dual part of its restriction to $M_4.$  

This system of equations is used in \cite{Witten:2011zz} to define knot invariants categorifying Jones polynomials.  A knot $K\subset W_3$ is introduced at $y=0$ for all $t:$ $\mathbb{R}\times K\subset \mathbb{R}\times W_3\times 0\subset \mathbb{R}\times W_3\times \mathbb{R}_+.$  To specify the conditions at $y=0$ one chooses a triplet $t_1, t_2, t_3$ of endomorphisms of $E|_{y=0}$ satisfying $[t_a, t_b]=\epsilon_{abc} t_c$ and some dreibein $e_j^a$ on $W_3$ defined by $e_j^a e_i^b \delta_{ab}=g_{ij},$ where  $g_{ij}dx^idx^j$ is the metric on $W_3.$   In terms of these, away from the knot the boundary conditions are 
\begin{equation}\label{Eq:NahmPole}
B=\frac{e_j^a t_a}{y} \omega_j+O(y),
\end{equation}
and as $y\rightarrow\infty$ one imposes $B\rightarrow0.$  The local condition near the knot can be formulated using the local coordinates aligned with it.  Orienting the knot along the $x^3$ direction and positioning it at the origin, one forms the complex  combination $B_{31}-\di B_{32}$.  The Haydys-Witten equations imply that this combination depends holomorphically on $z=x_1-\di x_2$ coordinate transversal to the knot.   This, in combination with condition \eqref{Eq:NahmPole} and the $B\rightarrow0$ condition at infinity, implies that locally $B_{31}-\di B_{32}$ has to have the form
\begin{equation}
B_{31}-\di B_{32}=\frac{1}{y}g^{-1}\left(\begin{array}{ccccc}
0& z^{r_1}&0&\ldots&0\\
0&0&z^{r_2}&\ldots&0\\
\vdots&\vdots&\vdots&\ddots&\vdots\\
0&0&0&\vdots &z^{r_\alpha}\\
0&0&0&\vdots &0
\end{array}  \right) g+O(y),
\end{equation}
with some gauge transformation $g.$  The set of integers $(r_1, r_2,\ldots, r_\alpha)$ corresponds to a representation assigned to the knot $K,$ that is part of the data determining the kind of the knot invariant that is being computed.

After this review we begin our exploration by writing the Haydys-Witten system of equations \eqref{Eq:HW} in octonionic form.

\section{Octonions}\label{Sec:Octonions}
\begin{figure}[htbp]
\begin{center}
\includegraphics[width=0.5\textwidth]{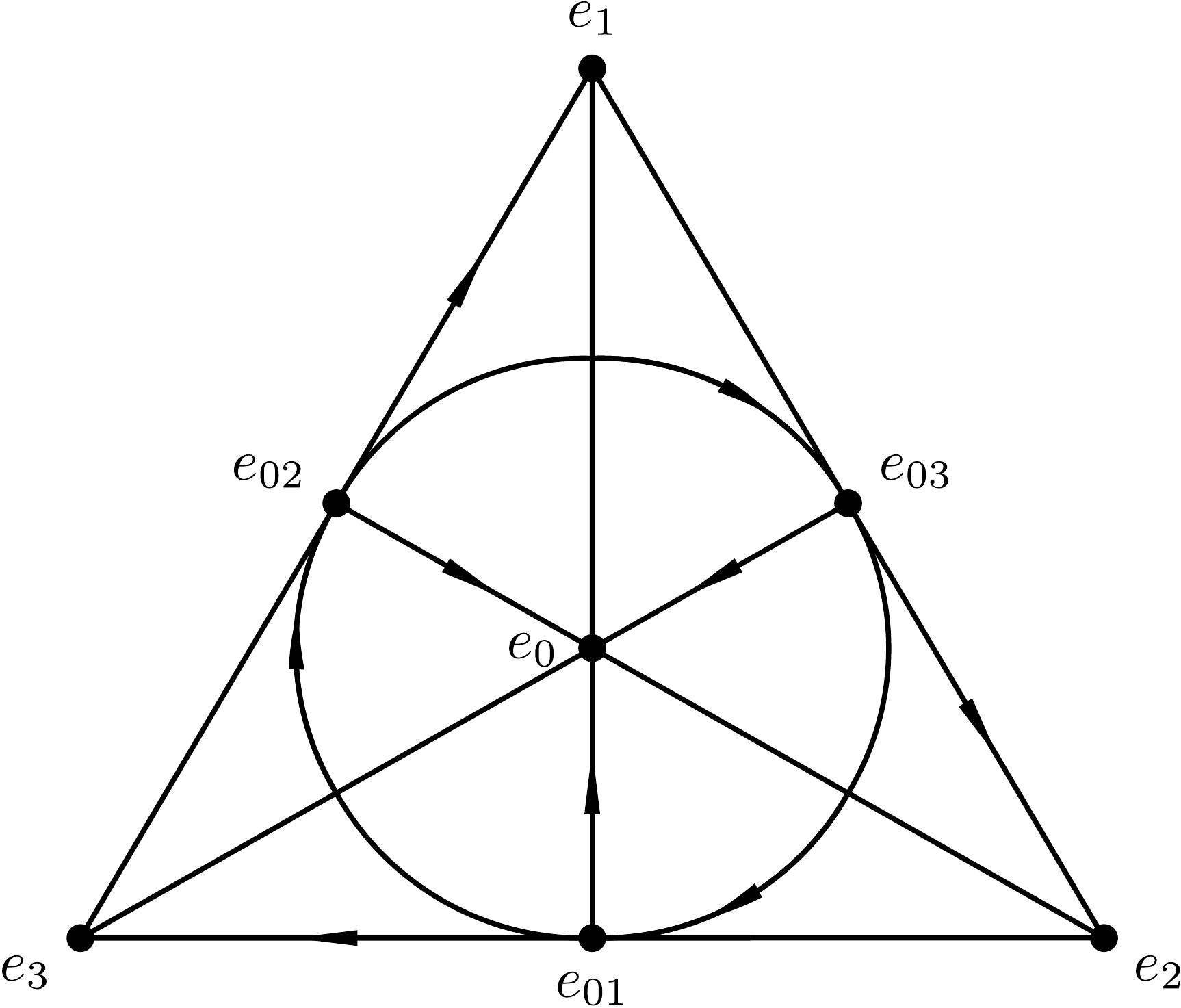}
\caption{Octonionic multiplication diagram.}
\label{UnitsTetragram}
\end{center}
\end{figure}
Octonions (or octaves) introduced by John Graves and Arthur Cayley form an eight-dimensional vector space over $\mathbb{R}$ spanned by $1, e_0, e_1, e_2, e_3, e_{01}, e_{02}, e_{03}.$  We use three ways to label the octonionic units: $(\hat{e}_{\hat{\mu}})=(e_\mu,e_{0j})=(e_0, e_j, e_{0j}),$ where $\hat{\mu}=0,1,\ldots,6,$ while $\mu=0,1,2,3$ and $j=1,2,3.$  Using these labelings (and implied summation over the repeated indices) an octonion $h$ can be written as 
\begin{equation}
h=y+q_0 e_0+ p_j e_{0j} +q_j e_j=y+q_\mu e_\mu+p_j e_{0j}=y+h_{\hat{\mu}} \hat{e}_{\hat{\mu}},
\end{equation}
and its conjugate is 
\begin{equation}
h^*=y-q_0 e_0- p_j e_{0j} -q_j e_j=y-q_\mu e_\mu-p_j e_{0j}=y-h_{\hat{\mu}} \hat{e}_{\hat{\mu}}.
\end{equation}
An octonion is a sum of its real and imaginary parts.  The real part of $h$ is defined to be ${\rm Re}\, h:=(h+h^*)/2=y$ and its imaginary part is ${\rm Im}\, h:=(h-h^*)/2=h-y.$

Octonions form an algebra with all octonionic units $e_0, e_j$ and $e_{0j}$ anticommuting with each other and each squaring to negative one:  $e_0^2=e_j^2=e_{0j}^2=-1.$ The rest of the multiplication rules are schematically summarized by Figure~\ref{UnitsTetragram}.  Every triplet of octonionic units lying on a line or a circle in this figure forms a triplet of quaternionic units. We note that $e_1, e_2, e_3$ do not form a triplet of quaternionic units, while $e_{01}, e_{02}, e_{03}$ do.  In particular
\begin{align}\label{Eq:Oct1}
e_{0j}e_{0k}&=-\delta_{jk}+\epsilon_{jkl}e_{0l},&
e_0 e_j&=e_{0j}=-\frac{1}{2}\epsilon_{jkl} e_k e_l,&
e_0 e_{0j}&=-e_j,\\
\label{Eq:Oct2}
e_j e_k&=-\delta_{jk}-\epsilon_{jkl}e_{0l},&
e_j e_{0k}&=\delta_{jk} e_0-\epsilon_{jkl}e_l.&
\end{align}
This defines the structure constants $f_{\hat{\mu}\hat{\nu}\hat{\rho}}$ such that $\hat{e}_{\hat{\mu}}\hat{e}_{\hat{\nu}}=-\delta_{\hat{\mu}\hat{\nu}}+f_{\hat{\mu}\hat{\nu}\hat{\rho}}\hat{e}_{\hat{\rho}}.$

\subsection*{Eight-dimensional Equation and its Reductions}\label{Sec:8dim}
Given a connection on a bundle $E\rightarrow\mathbb{R}^8,$ with some trivialization, we form a covariant differential 
$$dx^M\wedge D_M=dy\wedge D_y+dx^{\hat{\mu}}\wedge D_{\hat{\mu}}:=d+A\wedge,$$  consider a combination 
\begin{equation}\label{Eq:Dirac}
\D=D_y+\sum_{\hat{\mu}=1}^7 \hat{e}_{\hat{\mu}} D_{\hat{\mu}},
\end{equation}
and its  conjugate $\D^*=-D_y+\sum_{\hat{\mu}=1}^7 \hat{e}_{\hat{\mu}} D_{\hat{\mu}}.$  The `Laplacian' $\D^*\D$ is also an octonion with real and imaginary parts. Let us demand the `Laplacian' be octonionic real, i.e. 
\begin{equation}\label{Eq:General}
{\rm Im}\, \D^*\D=0.
\end{equation}
In flat space $\D^*\D=-D_y^2-D_{\hat{\mu}}D_{\hat{\mu}}+\hat{e}_{\hat{\rho}}\left(\frac{1}{2} f_{\hat{\mu}\hat{\nu}\hat{\rho}}[D_{\hat{\mu}}, D_{\hat{\nu}}]-[D_y, D_{\hat{\rho}}]\right).$ Thus we are led to the eight-dimensional equation 
\begin{equation}\label{Eq:Eight}
2 F_{y\hat{\rho}}=f_{\hat{\mu}\hat{\nu}\hat{\rho}} F_{\hat{\mu}\hat{\nu}}.
\end{equation}
It ensures that the `Laplacian' commutes with all octonionic units and equals to the negative of the covariant Laplacian.

One would like to view $\D$ as a linear operator acting on some Hilbert space.  
Since the division algebra of octonions is nonassociative, one might worry about the meaning of the Laplace operator $\D^* \D$ and of the above operator equation, as generally for a pair of two octonionic operators $A$ and $B$ we have $A(B \psi)\neq(AB)\psi.$  Octonions, however, do satisfy\footnote{The reason behind this associativity relation is that one can choose the basic octonions so that $A$ and $\psi$ lie in a three-space spanned by $1$ and two octonionic units.  Two octonionic units (if distinct) define a quaternionic line, containing $A$ and $\psi,$ and quaternions are associative.} $A^*(A\psi)=(A^*A)\psi.$   If one interprets $\D$ as an operator, Eq.~\eqref{Eq:General} implies that the operator $\D^* \D$ is purely real, moreover, it equals  the negative of the covariant Laplacian  which, for any nontrivial connection without flat factors, is strictly negative.  Thus there is a Green's function $G=(\D^* \D)^{-1}$ that, in turn, commutes with the octonionic units. 

The eight-dimensional equation \eqref{Eq:Eight} is not new; it first appears in \cite{CDFN,Ward}.  Some of its solutions (with gauge group SO(7) or SO(8)) are found in \cite{FairlieNuyts84} and \cite{FubiniNicolai85}.  Solutions with gauge group $SU(2)$ of the reduction of these equations to four-dimensional hyperk\"ahler base are analyzed in \cite{Dunajski:2011sx}.

\subsection*{Reduction to Five Dimensions}
Consider a reduction of the eight-dimensional equation \eqref{Eq:Eight} to five dimensions.  In other words, consider a connection with components that in some gauge are independent of three of the coordinates, say $x^{\hat{4}}, x^{\hat{5}}$ and $x^{\hat{6}}$.  In this gauge, denoting its components in these three directions  by $\Phi_1, \Phi_2,$ and $\Phi_3,$ we have
\begin{equation}
\label{Eq:HWDirac}
\D=D_y+e_{\mu}D_{\mu}+e_{0j}\Phi_j.
\end{equation}
Now the equation ${\rm Im}\, \D^*\D=0$ is the dimensional reduction of \eqref{Eq:General} and it reads
\begin{align}
&F_{y0}-D_j\Phi_j=0,\nonumber\\
\label{Eq:HWinR3}
&F_{yl}+D_0\Phi_l+\epsilon_{ljk}\left(D_j\Phi_k-D_k\Phi_j\right)=0,\\
&F_{l0}+\epsilon_{ljk}F_{jk}+D_y\Phi_l-\epsilon_{ljk}\Phi_j\Phi_k=0.\nonumber
\end{align}
At this point we observe that this is exactly the Haydys-Witten system of equations \eqref{Eq:HW} when $W_3=\mathbb{R}^3$ and $B=\Phi_j\left(\frac{1}{\sqrt{2}}dx^0\wedge dx^j+\frac{\epsilon^{jkl}}{2\sqrt{2}} dx^k\wedge dx^l\right).$

\subsection*{Reduction to Seven Dimensions}
Reducing the eight-dimensional equation to seven dimensions along the  $y$-direction leads to 
\begin{equation}\label{Eq:Octopole}
-D_{\hat{\rho}}\Phi=\frac{1}{2}f_{\hat{\mu}\hat{\nu}\hat{\rho}}F_{\hat{\mu}\hat{\nu}}.
\end{equation}
One cannot fail to notice its close resemblance of the Bogomolny equation. The role of the three-dimensional totally antisymmetric tensor of the Bogomolny equation is played here by the octonionic structure constants,  thus, we call Eq.~\eqref{Eq:Octopole} the {\em octonionic monopole equation}.

All of the equations appearing in this section are based on octonionic structure.  General manifolds respecting this structure are those with the G2 holonomy.  We give covariant formulation of these equations below in terms of the special holonomy structures.

\subsection*{Reduction to Three Dimensions}
The forthcoming brane picture of Section~\ref{Sec:Branes} suggests that a reduction of the eight-dimensional equations to three dimensions might lead to yet another description of  the knot invariants.  To achieve this reduction we let
$\D=Y+e_{0j}D_j+e_\mu \Phi_\mu.$  Now the condition ${\rm Im}\, \D^*\D=0$ reads
\begin{align}
\label{Eq:3Da}
&[D_j, \Phi_j]+[Y,\Phi_0]=0,\\
\label{Eq:3Db}
&[D_k,\Phi_0]+[\Phi_k,Y]-\epsilon_{ijk}[D_i, \Phi_j]=0,\\
\label{Eq:3Dc}
&[D_k,Y]-[\Phi_k,\Phi_0]+\frac{1}{2}\epsilon_{ijk}\left([D_i, D_j]-[\Phi_i,\Phi_j]\right)=0.
\end{align}
This is a three-dimensional system of partial differential equations on $\mathbb{R}_t\times\mathbb{R}_+\times S^1.$
To be useful for the knot invariants discussion, $(\Phi_2, \Phi_3, Y)$ should satisfy a Nahm triplet boundary condition at a codimension one boundary, while $\Phi_0,\ldots,\Phi_3$ form ADHM instanton data.
Just as ADHM equations, these equations should be slightly modified by introducing fundamental and antifundamental variables (respectively denoted by $I$ and $J$) that modify the right-hand-side.  The appropriate form of Eq.~\eqref{Eq:3Dc}  in the system (\ref{Eq:3Da},\ref{Eq:3Db},\ref{Eq:3Dc}) above appears to be 
\begin{align}
[D_1,Y]-[\Phi_1,\Phi_0]+\left([D_2, D_3]-[\Phi_2,\Phi_3]\right)&={\rm Re}\, IJ,\nonumber\\
\label{Eq:3DCorr}
[D_2,Y]-[\Phi_2,\Phi_0]+\left([D_3, D_1]-[\Phi_3,\Phi_1]\right)&={\rm Im}\, IJ,\\
[D_3,Y]-[\Phi_3,\Phi_0]+\left([D_1, D_2]-[\Phi_1,\Phi_2]\right)&=\frac{1}{2}(II^*-J^*J),\nonumber
\end{align}
augmented by appropriate equations involving covariant derivatives of $I$ and $J.$
We leave this three-dimensional discussion, however interesting, outside of the scope of this note, emphasizing  instead the relation between the five- and the seven-dimensional systems of equations.

\section{Covariant Forms of the Octonionic Equations}\label{Sec:Covariant}
So far our focus is on differential equations in flat spaces induced by the octonionic structure.  Our next goal is to put these equations in a covariant form, so that they generalize to a wider range of backgrounds.  To begin with, we collect some relevant facts involving special holonomy.  A wealth of information on special holonomy and the relation of octonions to G2 and Spin(7) structures can be found in \cite{Salamon:2010cs}.

\subsection*{Special Holonomy}\label{Sec:SpecialHol}
The cotangent space $T^*W_3$ of a three-manifold $W_3$ can be given a Calabi-Yau structure in the vicinity of its zero section.  We would strongly prefer to work with $W_3$ such that its cotangent bundle  $T^*W_3$ is a complete Calabi-Yau manifold.   For which $W_3$ this is the case appears to be an open problem.  Besides the flat three-space $W_3=\mathbb{R}^3$ and flat three-tori $W_3=T^3,$ another example of a three-manifold satisfying this condition is the three-sphere $W_3=S^3.$ The former two are the basic examples we have in mind in what follows.  In the case of $W_3=\mathbb{R}^3$ with coordinates $q_1, q_2,$ and $q_3,$ and conjugate coordinates $p_1, p_2,$ and $p_3$ in the fiber of $T^*W_3,$ then the symplectic two-form is $\omega=dq_j\wedge dp_j,$ and the holomorphic three-form is $\eta=d(p_1+i q_1)\wedge d(p_2+i q_2)\wedge d(p_3+i q_3),$ so that $T^*W_3=T^*\mathbb{R}^3=\mathbb{R}^6$ is a complete Calabi-Yau (in this case, the flat six-space).  
For any Calabi-Yau space $(CY,\omega,\eta)$ there is a G2 structure on $\mathbb{R}_t\times CY$ given by the three-form and the four-form
\begin{align}
\phi^{(3)}&={\rm Re}\,\left( e^{i\alpha}\eta\right)+\omega\wedge dt,\\
\psi^{(4)}&=*\phi^{(3)}=\frac{1}{2}\omega\wedge\omega+{\rm Im}\,\left( e^{i\alpha} \eta\right)\wedge dt,
\end{align}
with any chosen real constant $\alpha.$ 
The group of linear transformations leaving the three-form $\phi^{(3)}$ invariant is G2.  This form can locally be put into a canonical form $\phi^{(3)}=\frac{1}{6}f_{\hat{\mu}\hat{\nu}\hat{\rho}}dx^{\hat{\mu}}dx^{\hat{\nu}}dx^{\hat{\rho}},$ where $f_{\hat{\mu}\hat{\nu}\hat{\rho}}$ are the octonionic structure constants;  thus G2 can be viewed as the group of symmetries of imaginary octonions.  A particular example of a manifold with G2 holonomy relevant to our discussion is $\mathbb{R}_t\times T^*W_3.$

In turn, for any G2 manifold $(Y^7, \phi^{(3)}, \psi^{(4)})$ there is a natural Spin(7) structure on $Y^7\times\mathbb{R}$ given by the four-form $\Omega^{(4)}=dy\wedge\phi^{(3)}+\psi^{(4)}.$  The relevant Spin(7) holonomy manifold for our discussion is $\mathbb{R}_t\times T^*W_3\times \mathbb{R}_+.$

As our equations involve the curvature form algebraically, two-forms on these manifolds are particularly important, and now we describe their structure.

\subsubsection*{G2 Structure}
Since G2 acts on the seven-dimensional space $V$ of imaginary octonions, it acts on the two-forms $\Lambda^2V^*.$ This 21-dimensional representation of G2 splits as a sum of the seven- and the  fourteen-dimensional irreducible representations (see  \cite{Salamon:2010cs} and \cite{ReyesCarrion:1998si}): $\Lambda^2V^*=\Lambda^2_7\oplus \Lambda^2_{21}.$ These irreducible representations have the following characterization \cite{Salamon:2010cs} in terms of the G2 three- and four-forms:
\begin{align}
\Lambda^2_7&=\left\{\alpha\in\Lambda^2V^* \Big| *(\phi^{(3)}\wedge\alpha)=2\alpha  \right\}
=\left\{\imath_{v}\phi^{(3)} \Big| v\in V\right\}\nonumber \\
 &=\left\{\alpha\in\Lambda^2V^* \Big| *(\psi^{(4)}\wedge*(\psi^{(4)}\wedge\alpha))=3\alpha  \right\},\label{Eq:Seven}\\
 \Lambda^2_{14}&=\left\{\alpha\in\Lambda^2V^* \Big| *(\phi^{(3)}\wedge\alpha)=-\alpha  \right\}
 =\left\{\alpha\in\Lambda^2V^* \Big| \psi^{(4)}\wedge\alpha=0  \right\},
 \label{Eq:Fourteen}
\end{align}
These characterizations allow us to write the projection operator onto $\Lambda_7^2$ space $\pi_7:\Lambda^2V^*\rightarrow \Lambda^2_7$  in two ways:
\begin{equation}\label{Eq:Proj}
\alpha\mapsto\pi_7(\alpha)=\frac{1}{3}(\alpha+*(\phi^{(3)}\wedge\alpha))=\frac{1}{3}*(\psi^{(4)}\wedge*(\psi^{(4)}\wedge\alpha)).
\end{equation}

On a G2 manifold the three-form $\phi^{(3)}$ gives an associative calibration and the four-form $\psi^{(4)}$ gives a coassociative calibration. A three-dimensional  submanifold $A^3_\Sigma\subset Y^7$ is called associative  if the restriction of the three-form $\phi^{(3)}$ to it is the volume form of $A^3_\Sigma:$  $\phi^{(3)}|_{A^3_\Sigma}={\rm Vol}_{A^3_\Sigma}.$
A four-dimensional submanifold $C^4_W\subset Y^7$ is coassociative if the restriction of the four-form $\psi^{(4)}$ to it equals its volume form: $\phi^{(4)}|_{C^4_W}={\rm Vol}_{C^4_W}.$

\subsubsection*{Spin(7) Structure}
Similarly, Spin(7) group action on an eight-dimensional space $U$ induces an action on its dual $U^*$ and on $\Lambda^2 U^*,$ which splits into irreducible representations: 
$\Lambda^2 U^*=\Lambda^2_7\oplus\Lambda^2_{21}.$  The two factors can be characterized by  
\begin{align}
\label{Eq:Dec1}
 \Lambda^2_7&=\left\{\alpha\in\Lambda^2U^* \Big| *(\Omega^{(4)}\wedge\alpha)=3\alpha  \right\},\\
 \label{Eq:Dec2}
 \Lambda^2_{21}&=\left\{\alpha\in\Lambda^2U^* \Big| *(\Omega^{(4)}\wedge\alpha)=-\alpha  \right\},
\end{align}

\subsection*{The Spin(7) Instanton and its Reductions}\label{Sec:Spin7}
Given any G2 manifold $(Y^7, \phi^{(3)}, \psi^{(4)})$ consider a Spin(7) manifold $\mathbb{R}_y\times Y^7$ with the Spin(7) structure given by $\Omega^{(4)}=dy\wedge\phi^{(3)}+\psi^{(4)}.$  The covariant form of eight-dimensional Eq.~\eqref{Eq:Eight} is
\begin{equation}
\imath_{\frac{\partial}{\partial y}}F_{\mathbb{A}}=*_7(\psi^{(4)}\wedge F_{\mathbb{A}}).
\end{equation}
In fact, this equation is equivalent to the Spin(7) instanton  equation of \cite{CDFN,Ward,ReyesCarrion:1998si,Donaldson:1996kp} that makes sense on any Spin(7) manifold and reads
\begin{equation}\label{Eq:Spin7}
*_8(\Omega^{(4)}\wedge F_{\mathbb{A}})=-F_{\mathbb{A}}.
\end{equation}
According to the decomposition $\Lambda^2U^*=\Lambda^2_7\oplus\Lambda^2_{21}$ and Eqs.~\eqref{Eq:Dec1} and \eqref{Eq:Dec2}, it states that the curvature two-form $F$ has no $\Lambda_7^2$ components.

This Spin(7) instanton equation has an important geometric interpretation \cite{ReyesCarrion:1998si,Donaldson:1996kp} as a gradient flow of the Chern-Simons functional on the space of connections on a G2 manifold $Y^7:$
\begin{equation}
{\rm CS}^\psi=\frac{1}{2}\int_{Y^7} {\rm tr}\, \left(A\wedge dA+\frac{2}{3} A\wedge A\wedge A\right) \wedge \psi^{(4)}.
\end{equation}

\subsubsection*{Dirac Operator Interpretation and Integrability}
Throughout Section~\ref{Sec:Octonions} we use $\D$ as a formal combination in order to use the octonionic structure to encode either the Spin(7) instanton, or a solution of the Haydys-Witten equations,  or an octonionic monopole.  We would like, however, to use $\D$ as an operator and its Hermitian conjugate be $\D^*$.  In order to do that we have to specify the Hilbert  space on which it acts.  Following \cite{Salamon:2010cs}, given a G2 manifold, we consider the spin bundles $S^+=\Omega^0\oplus\Omega^2_7$ and $S^-=\Omega^1.$  Both are rank $8$ and can be identified with octonions.  Left multiplication gives the action of imaginary octonions on $S^+:$ for $a^*=-a$ let $L_a:v\mapsto av.$ Then $L_a^2 v=a(av)=-a^*(av)=-(a^*a)v$ and  $L_a^2=-|a|^2.$ Thus  the action of imaginary octonions on $S^+$   by the left multiplication forms an eight-dimensional representation of the Clifford algebra in seven dimensions.  As a result, we can view the operator $\D$ in seven dimensions (and its reduction to five dimensions used above) as a Dirac operator. 

The Clifford multiplication is given \cite{Salamon:2010cs} by
\begin{align}
&\begin{array}{rccl}
\gamma(e):&S^+&\rightarrow& S^-\\
&(f,\omega)&\mapsto & f e^*+2\imath_e\omega \end{array},
\end{align}
and
\begin{align}
&\begin{array}{rccl}\gamma(e)^*:&S^-&\rightarrow& S^+\\
&v&\mapsto& ((e,v), \frac{1}{2}(\imath_e\imath_v\Omega^{(4)}+e^*\wedge v^*)), \end{array}
\end{align}
which we write in  Eq.~\eqref{Eq:Gammas} below in component form.

In fact, the Spin(7) instanton equation and all of its reductions can be interpreted as an integrability condition for a linear system.  This linear system is exactly $\D^*\Psi=0.$ This fact, though in a different form, was discovered in \cite{Fairlie:1984bh}.  Let $\Psi=\hat{e}_{\hat{\rho}}\Psi^{\hat{\rho}}$, while $\D^*=-D_0+\hat{e}_{\hat{\rho}}D_{\hat{\rho}}$ as defined in Sec.~\ref{Sec:8dim}, then the components of the equation $\D^*\Psi=0$ form a system of eight equations.  Let us begin by writing it in component form by introducing a tensor\footnote{This is an eight-dimensional analogue of the 't Hooft tensor in four dimensions.} $R^{\hat{\rho}}_{MN}$ (antisymmetric in $MN$) with $\hat{\rho}=1,\ldots,7$ and $M,N=0,1,\ldots,7$ which is defined by the following relation
\begin{equation}
\hat{e}_{\hat{\rho}}R^{\hat{\rho}}_{MN} dx^M\wedge dx^N=dx^*\wedge dx,
\end{equation}
where $dx=dx^0+\hat{e}_{\hat{\rho}}dx^{\hat{\rho}}$ and $dx^*=dx^0-\hat{e}_{\hat{\rho}}dx^{\hat{\rho}}.$  The components of this tensor are exactly the Clebsch-Gordan coefficients or $\underline{8}\wedge\underline{8}\rightarrow\underline{7}$ in the decomposition $\Lambda^2U^*=\Lambda^2_7\oplus\Lambda^2_{21}$ of Eqs.~(\ref{Eq:Dec1},\ref{Eq:Dec2}), which form the representation decomposition  $\underline{8}\wedge\underline{8}=\underline{7}+\underline{21}.$  Then, our key equation $\D^*\Psi=0$ in components reads
\begin{equation}\label{Eq:R}
R^{\hat{\rho}}_{MN}D_M\Psi^{\hat{\rho}}=0,
\end{equation}
and the integrability condition for this system of linear equations is 
\begin{equation}
R^{\hat{\rho}}_{MN} F_{MN}=0.
\end{equation}
This means that in the $\Lambda^2=\Lambda^2_7\oplus\Lambda^2_{21}$ decomposition of the curvature two-form $F=F_7+F_{21}$ the $F_7\in\Lambda^2_7$ component vanishes. This is exactly the meaning of the Spin(7) instanton condition \eqref{Eq:Spin7}.  

This also gives a direct interpretation of the operator $\D^*$ as a chiral Dirac (aka Weyl) operator.  Namely, one can represent gamma matrices in eight dimensions by
\begin{align}
\label{Eq:Gammas}
\gamma^0&=\left(\begin{matrix} 
0 & 1 \\ 
1 & 0
\end{matrix}\right),&
\gamma^{\hat{\rho}}&=\left(\begin{matrix} 
0 & R^{\hat{\rho}} \\ 
-R^{\hat{\rho}} & 0
\end{matrix}\right),
\end{align}
with the off-diagonal $8\times 8$ matrices $R^{\hat{\rho}}=(R^{\hat{\rho}}_{MN}).$ Such a relation between spinors and octonions is pointed out explicitly in \cite{CDFN} for example.  Thus the linear equation \eqref{Eq:R} acquires an interpretation as an eight-dimensional Dirac (or, rather, Weyl) equation.  

\section{Octonionic Monopole}
At this point we possess the ingredients needed to formulate the covariant system in seven-dimensions that we conjecture to be dual to the Haydys-Witten five-dimensional system.  It is based on the seven-dimensional  equation \eqref{Eq:Octopole}.  We proceed by writing it in a covariant form.

On any G2 manifold $Y^7$ there is a structure of imaginary octonions on each tangent space with the structure constants given by the components of the three-form $\phi^{(3)}$.  For a Hermitian vector  bundle $\hat{E}\rightarrow Y^7$ with a connection $A$ and an endomorphism $\Phi$ we write a covariant form of the octonionic monopole Eq.~\eqref{Eq:Octopole} as
\begin{equation}\label{Eq:CovMon} 
*\left(\psi^{(4)}\wedge F_A\right)=-D\Phi.
\end{equation}

We continue the parallel with monopoles, where the conventional Bogomolny equation in three dimensions is a reduction of the four-dimensional self-duality equation.  The octonionic monopole equation above on a G2 manifold $Y^7$, in turn, can be viewed as a reduction of the Spin(7) instanton equation \eqref{Eq:Spin7} on $\mathbb{R}_y\times Y^7.$ 
That is how it appears in \cite[Eq.(25)]{Donaldson:2009yq}.  
 Namely,  if the connection one-form  $\mathbb{A}= \Phi dy+A$ on $\mathbb{R}_y\times Y^7$ has components independent of the variable $y$ then its curvature  two-form is $F_\mathbb{A}=dy\wedge D\Phi+F$ and the Spin(7) instanton equation \eqref{Eq:Spin7} amounts to the following system of two equations
\begin{align}\label{Eq:OctMon1} 
*_7(\psi^{(4)}\wedge F)+D\Phi=0,\\
\label{Eq:OctMon2}
*_7 F+\phi^{(3)}\wedge F+\psi^{(4)}\wedge D\Phi=0.
\end{align}
These two, however, are equivalent thanks to relation \eqref{Eq:Proj}.  Thus we have two equivalent forms of the octonionic monopole equation \eqref{Eq:OctMon1} and \eqref{Eq:OctMon2}.  Geometrically it  states that the $\Lambda_7^2$ component of the curvature two-form $F$ is given in terms of the gradient of the Higgs field $\Phi.$

The octonionic monopole equation (modulo gauge transformations) is  elliptic.  Its linearization is the following elliptic complex
\begin{align}
{\rm ad} (E)\otimes\Omega^0\xrightarrow{\delta_0}
{\rm ad} (E)\otimes(\Omega^0\oplus\Omega^1)\xrightarrow{\delta_1}
{\rm ad} (E)\otimes\Omega^1,
\end{align}
with the differentials
\begin{align}
\delta_0:& \lambda\mapsto \left(
   \begin{matrix} 
      [\Phi, \lambda] \\
      D_A\lambda \\
   \end{matrix}
   \right),&
\delta_1:& \left(   \begin{matrix} 
      \alpha \\
      \beta \\
   \end{matrix}
   \right)\mapsto D_A\alpha-[\Phi, \beta]+*(\psi^{(4)}\wedge D_A\beta).
\end{align}
For any application to knot invariants a careful study of the ellipticity of its boundary conditions formulated below is needed.

\subsection*{Energy Bound}
The topological charges\footnote{These charges are topological in a limited sense, as they are invariant under metric deformations that preserve the special holonomy structure.} one can associate to appropriate solutions of Eq.~\eqref{Eq:CovMon} are an instanton number $I=\int_{Y^7} \phi^{(3)}\wedge{\rm tr}\, (F\wedge F),$ a monopole number $M=\int_{Y^7} \psi^{(4)}\wedge{\rm tr}\left(D\Phi\wedge F\right),$ and an octopole charge $O=\int_{Y^7} {\rm tr}\, F\wedge F\wedge F\wedge D\Phi.$

A natural action functional assigned to any pair $(A,\Phi)$ is
\begin{equation}
S[A,\Phi]=-\int_{Y^7} {\rm tr}\, \left(F\wedge *F+D\Phi\wedge *D\Phi\right).
\end{equation}
The fields in our conventions are antihermitian, hence the minus sign in this definition. 
Using the identity $*(\psi^{(4)}\wedge \alpha)\wedge\psi^{(4)}\wedge\alpha=\alpha\wedge*\alpha+\alpha\wedge\alpha\wedge\phi^{(3)}$ valid for any two-form $\alpha,$ we obtain the inequality
$0\geq\int_{Y^7}{\rm tr}\left(D\Phi+*(\psi^{(4)}\wedge F)\right)\wedge*\left(D\Phi+*(\psi^{(4)}\wedge F)\right)=I+2M-S.$  Thus we obtain the Bogomolny-type inequality
\begin{equation}
S\geq I+2M.
\end{equation}
This inequality is saturated only by octonionic monopoles.  Thus octonionic monopoles are solutions of the seven-dimensional Yang-Mills-Higgs equations that deliver the action minimum in any given topological class.  This fact also follows from the octonionic monopole Eq.~\eqref{Eq:CovMon} and the Bianchi identity.
 
In a noncompact case one would rewrite the instanton and monopole charges in terms of boundary integrals using 
$$I=\int_{\partial Y^7} \phi^{(3)}\wedge{\rm tr}\, \left(A\wedge dA+\frac{2}{3}A\wedge A\wedge A\right),$$ and  $M=\int_{Y^7} \psi^{(4)}\wedge{\rm tr}\,\left( D\Phi\wedge F\right)=\int_{\partial Y^7} \psi^{(4)}\wedge{\rm tr}\, \left(\Phi\wedge F\right).$  

\subsection*{A Knot at the Heart of a Monopole}
In a situation dual to Witten's picture of Section~\ref{HaydysWitten}, given a knot $K\subset W_3$ at $y=0$  we consider the manifold $Y^7=\mathbb{R}_t\times T^*W_3$ as a G2 manifold with a coassociative cycle $C^4_W=\mathbb{R}_t\times W_3$ at the zero section of $T^*W_3.$ Imagine, for the time being, that we also have an associative noncompact cycle $A^3_\Sigma=\mathbb{R}_t\times\Sigma$ with a boundary in $C^4_W$ specified by the knot: $\partial \Sigma=K\subset W_3\subset T^*W_3.$  One can consider a more general situation with any G2 manifold $Y^7$ and a pair of cycles $C^4_W$ and $A^3_\Sigma,$ such that $C^4_W$ is coassociative and $A^3_\Sigma$ is an associative cycle\footnote{The associative condition in fact is only needed at the boundary $\partial A^3_\Sigma.$} with some fixed boundary $\partial A^3_\Sigma \subset C^4_W.$

We would like to impose boundary conditions  corresponding to a situation with a monopole at  $C^4_W$ and an instanton positioned at $A^3_\Sigma$.  Loosely speaking this signifies the following.
The asymptotic boundary of $T^*W_3$ is a sphere bundle over $W_3.$ We demand that this sphere fiber has a monopole flux through it contributing to the monopole charge  $M.$ The knot, on the other hand, is the boundary of the  submanifold $A^3_\Sigma$ of codimension four.  Moving along $A^3_\Sigma$ and away from $W_3,$ we demand that there is a Yang-Mills instanton of a given instanton number $k$ on the four-space transverse to $A^3_\Sigma.$ Our motivation for this choice, together with the local conditions near the knot, stems from the brane configuration described in the next section.  

While the zero section of $T^*W_3$ is always present to produce the coassociative cycle $C^4_W=\mathbb{R}_t\times W_3,$ given a knot $K\subset W_3,$ there might not exist an appropriate associative cycle $A_\Sigma^3$ ending on it (as described above).  In terms of the octonionic monopole, however, the knot is encoded in the boundary conditions at $C^4_W$ and the remnant of (nonexistent) associative cycle $A_\Sigma^3$ is the condition of nonzero instanton number.

\section{Branes and Probes}\label{Sec:Branes}
Witten's description \cite{Witten:2011zz} of Khovanov homology for a knot $K$ in $W_3$ can be viewed as arising from the following configuration.  One studies M theory on $\mathbb{R}_t\times T^*W_3\times TN$ with $n$ M5-branes with world-volumes $\mathbb{R}_t\times W_3\times C,$ here $W_3$ is the zero section of $T^*W_3$ and $C$ is the `cigar' semi-infinite cycle of the Taub-NUT space $TN.$  For a knot $K\subset W_3$ one introduces an M2-brane with the world-volume $\mathbb{R}_t\times \Sigma_K\approx\mathbb{R}_t\times\mathbb{R}_+\times S^1$ with boundary   $\mathbb{R}_t\times K$.  The M2-brane is ending on the M5-branes and its boundary is the knot positioned in $W_3$ at the tip of the `cigar' $C$. We view $T^*W_3$ as a Calabi-Yau space  and require supersymmetry of this brane configuration.

Identifying the circle of the Taub-NUT as the circle of M theory we obtain an equivalent Type IIA  brane configuration:
\begin{align}\label{Diag:IIA}
     &\mathbb{R}_t\times T^*W_3\times \mathbb{R}^3\nonumber\\
D6\ \ &\mathbb{R}_t\times T^*W_3\times 0\\
D4\ \ &\mathbb{R}_t\times \phantom{T^*}W_3\times \mathbb{R}_+ \nonumber\\
D2\ \ &\mathbb{R}_t\times\phantom{T} \Sigma_K \nonumber
\end{align}

The five-dimensional equations \eqref{Eq:HW} are used in  \cite{Witten:2011zz} to describe this configuration in terms of the theory on the world-volume of the D4-brane.  Here we attempt to obtain an equivalent description in terms of the theory on the world-volume of the D6-branes.  (One might also attempt to give a description in terms of the theory on the D2-brane.)  Such a description would involve Eqs.~\eqref{Eq:3DCorr}.  The fundamental and antifundamental degrees of freedom, respectively I and J, in these equations correspond to open strings connecting D2- to D4-branes.) The seven-dimensional system of equations that ensures the same supersymmetry condition on the D6 is to be given by the octonionic monopole Eq.~\eqref{Eq:CovMon}.  If the  Haydys-Witten equations \eqref{Eq:HW} on the D4-branes  were for rank $n$, with residue triplet $(t_1,t_2,t_3)$ of the boundary condition \eqref{Eq:NahmPole} at $y=0$ consisting of $L$ irreducible representations of $su(2)$, then the dual octonionic monopole equations \eqref{Eq:CovMon} on D6-branes are for rank $L$ connection and Higgs fields such that 1) their limiting behavior at the zero section of $T^*W_3$ is that of charge $n$ Dirac monopole (corresponding to $n$ semi-infinite D4-branes ending on the D6) and 2) they represent an instanton at $\Sigma_K.$  
The Nahm boundary conditions of \cite{Witten:2011pz} that are analyzed in \cite{Mazzeo:2013zga} have the irreducible Nahm pole, i.e. $L=1.$ 
From the seven-dimensional point of view this is the rank one case; it appears degenerate and might not capture all of the knot information we might want. It also suffers from the presence of zero size instantons.  One could introduce noncommutativity to regularize this configuration.  A better direction to pursue, however, is to increase the rank of the gauge group of the seven-dimensional theory.  This would amount to considering either a non-maximal or reducible pole in the Witten's picture or the five-dimensional theory on a finite $y-$interval, so that the base space is $\mathbb{R}_t\times W_3\times I.$  In general, the rank of the seven-dimensional system (the number of the D6-branes) is the number of irreducible representations of the residue of the Nahm pole in the boundary conditions of the five-dimensional system.  While the monopole charge (the number of D4-branes) carried by the coassociative cycle $\mathbb{R}_t\times W_3$ is the rank of the five-dimensional system.

\subsection*{The Model Solution}
In coordinate form the seven-dimensional octonionic monopole equation amounts to the following system
\begin{align}
D_{q_0}\Phi&=-\sum_{j=1}^3 F_{q_j p_j},\\
D_{q_i}\Phi&=F_{q_0p_i}+F_{q_j p_k}+F_{p_j q_k},\\
D_{p_i}\Phi&=-F_{q_0 q_i}+F_{q_j q_k}-F_{p_j p_k},
\end{align}
where we understand the triplet $(i,j,k)$ to be any cyclic permutation of $(1,2,3).$ 
Orienting the knot at a given point along the $q_3$-axis we are seeking  a model solution that is static and independent of $q_3.$  To begin with, we put $D_{q_0}=0$ and $D_{q_3}=0$ and introduce combinations $D_1=D_{q_1}-\di D_{q_2}, D_2=D_{p_1}+\di D_{p_2}$ and $D_3=D_{p_3}+\di\Phi.$  In terms of these coordinates the above equations take the form
\begin{align}
\label{Eq:Complex}
&[D_j, D_k]=0,\ {\rm for}\ j,k=1,2,3,
\\
\label{Eq:Real}
&[D_1,\bar{D}_1]+[D_2,\bar{D}_2]+[D_3,\bar{D}_3]=0.
\end{align}
For a fixed value of $p_3$ the equation $[D_1,D_2]=0$ implied that a solution gives a holomorphic bundle on each four-dimensional slice $p_3=const$ with holomorphic coordinates $q=q_1-\di q_2$ and $p=p_1+\di p_2.$  Moreover, the remaining two equations of \eqref{Eq:Complex} are $[D_3,D_1]=0$ and $[D_3,D_2]=0.$ They  imply that as we vary $p_3$ this holomorphic bundle does not change so long as we do not encounter  singularities.  We fix the instanton number of this bundle at some $p_3>0.$

We are seeking a solution with a Dirac monopole singularity at $p_3=p=0.$  It corresponds to the semi-infinite D4-branes ending on the D6-brane. So we expect a holomorphic bundle $E^+$ for any value of $p_3>0$ to be independent of $p_3$ and a bundle $E^-$ for any value of $p_3<0$ to be independent of the value of $p_3$ as well.  At $p_3=0$ we encounter a singularity only on the plane $p=0.$ Everywhere outside this plane $E^+=E^-.$  The singularity at $p_3=p=0$ is the Dirac singularity corresponding to the Hecke modification along $p=0$ relating $E^-$ to $E^+.$  The local analysis parallels exactly that of \cite{Kapustin:2006pk}.

The system (\ref{Eq:Complex},\ref{Eq:Real}) is a reduction to five dimensions of the Hermitian Yang-Mills equations.  According to the results of Donaldson and of Uhlenbeck and Yau,  Eq.~\eqref{Eq:Real} has a unique solution if the holomorphic bundle given by a solution of Eq.~\eqref{Eq:Complex} is semi-stable. In the present case such analysis is needed in the presence of the Dirac monopole at $p=0, p_3=0.$

\section{Conclusions and Speculations}
We gave the octonionic interpretation of Haydys-Witten equations \eqref{Eq:HW}.  This allows us to view them, as well as the octonionic monopole equation \eqref{Eq:CovMon}, as a reduction of Spin(7) instanton \eqref{Eq:Spin7}.  From the string theory brane configuration point of view \eqref{Diag:IIA}, we expect a relation between solutions of the Haydys-Witten system and octonionic monopoles; the former emerging in the world-volume of the D4-branes, while the latter on the world-volume of the D6-branes.   This relation is similar to the Nahm transform that relates two reductions of the self-dual Yang-Mills equation in four dimensions: solutions of the Nahm equation and solutions of the monopole equation of Bogomolny.  

A natural question to ask is whether there is a quantum theory that could lead to the octonionic monopole description. In fact there is a theory which has `topological' observables for the case we consider.  In \cite{Acharya:1997gp} the super-Yang-Mills (with no twisting) on a manifold with G2 holonomy  is argued to be a topological theory.  Namely, for $Q$ the BRST generator, the super-Yang-Mills Lagrangian is written in a  $Q$-exact form: $L=-\di\{Q,V\}$ and the BRST invariant configurations are identified as Spin(7) instantons.  Thus, following the reasoning of \cite{Witten:1988ze}, any correlation function of BRST closed operators is invariant under holonomy-preserving metric variations.  The right observable to consider in the case at hand has the operator insertion of the (codimension three) 't Hooft operator at the zero section of $T^*W_3$ together with a relative codimension two operator at the position of the knot $K\subset W_3$ within it, as described by the model solution of Section \ref{Sec:Branes}.

In the Donaldson-Segal picture \cite{Donaldson:2009yq}, the Spin(7) instanton equations on $Y^7\times\mathbb{R}$ emerge as the covariant form of the gradient flow for the Chern-Simons functional
\begin{equation}
{\rm CS}^\psi=\frac{1}{2}\int_{Y^7} {\rm tr}\, (A\wedge dA+\frac{2}{3}A\wedge A\wedge A)\wedge\psi^{(4)}.
\end{equation}
Similarly, the octonionic monopole equation on $CY\times\mathbb{R}$ can be interpreted as the gradient flow for the functional
\begin{equation}\label{Eq:fnl}
h(\Phi,A)=\frac{1}{2} \int_{CY} {\rm tr}\, (\Phi F) \wedge\omega^{(2)}\wedge\omega^{(2)}
+{\rm tr}\, (A\wedge dA+\frac{2}{3} A\wedge A\wedge A)\wedge {\rm Re}\, e^{i \alpha}\eta^{(3)}.
\end{equation}
Here $\alpha\in[0,2\pi)$ is some fixed arbitrary constant.
A fixed point of this gradient flow has $\delta h=0$ and satisfies
\begin{align}\label{Eq:fixed}
F\wedge\omega^{(2)}\wedge\omega^{(2)}&=0,\\
D\Phi\wedge\omega^{(2)}\wedge\omega^{(2)}&=2 F\wedge {\rm Re}\, e^{i\alpha}\eta^{(3)}.\label{Eq:fixed2}
\end{align}
The first equation involves the $(1,1)$ part of the curvature and can be viewed as a moment map condition, while the second equation is a holomorphic condition relating the $(0,2)$ component of the curvature to the holomorphic covariant differential of the Higgs field.

One can compare to Witten's five-dimensional generalization of the Floer-Donaldson theory. In order to obtain knot invariants, according to \cite{Witten:2011pz}, one  considers the Chern-Simons functional, or rather its imaginary part, on the space of  complexified connections on $W_3$ as a Morse function.  The complexified gauge group is $G^{\mathbb{C}}.$ Trying to formulate a gradient flow for such a function one faces a problem:  the metric is $G$ invariant, but not  $G^{\mathbb{C}}$ invariant.  An invariant metric is needed to obtain the flow vector field from the differential of the Morse functional.  This difficulty is solved by imposing  a moment map constraint $\mu=0$ and flowing on the level-set of this constraint.  The constraint condition  is enforced by introducing  a Higgs field Lagrange multiplier $\phi$ and modifying the Morse function to be $f=\int{\rm tr}\phi\mu+{\rm Im}\, CS^{G^{\mathbb{C}}}.$  The gradient flow equation on $W_3\times\mathbb{R}_+$ is the Kapustin-Witten equation.  It can be used to reproduce the Jones polynomials \cite{Gaiotto:2011nm}, which are `classical invariants' of the knot. 
This picture provides much insight by naturally reproducing Khovanov homology and by giving instanton interpretation to the integer (difference of)  powers of the terms in the Jones polynomial \cite{Witten:2014xwa}. 
  In order to introduce Khovanov homology, which encodes `quantum invariants' of the knot, one introduces an extra dimension and promotes the Lagrangian multiplier Higgs field $\phi$ to the connection component in that direction. The Hilbert space of this five-dimensional theory is the Khovanov homology.  It is described semiclassically. The resulting equations are the Haydys-Witten equations.  This is the view presented in  \cite{Witten:2011pz} and \cite{Witten:2011zz}.

In the seven-dimensional case, the picture appears to be more conventional and  in tune with Witten's version of Morse theory \cite{Witten:1982im}. One begins by considering the Chern-Simons-Higgs functional on the space of pairs $(A,\Phi)$ of a connection and an endomorphism of a hermitian bundle over a Calabi-Yau, such as $T^*W_3$, given by Eq.~\eqref{Eq:fnl}. It is purely real.  Its fixed points (solutions of Eqs.~(\ref{Eq:fixed},\ref{Eq:fixed2})) are used to span vector spaces of the Morse complex; while the gradient flow between them  (given by octonionic monopole solutions on $CY\times\mathbb{R}$),  interpolating between one fixed point solution on the Calabi-Yau at $t=-\infty$ and another at $t=+\infty,$ are to be  used to define the differential.  If, indeed, the correspondence  argued in Section~\ref{Sec:Branes} between octonionic monopoles and solutions of the Haydys-Witten equations holds, then the homology groups of this complex should reproduce the Khovanov homology.

\section*{ Acknowledgments}
This work was partially supported by a grant from the Simons Foundation (\#245643 to SCh).   
It is a pleasure to thank Michael Atiyah, Benoit Charbonneau, and Nigel Hitchin for stimulating discussions.  The author is grateful to Edward Witten for early explanations of his work on the subject.  A special thank you is due to  the organizers of the 2012 Banff workshop ``Special Holonomy and their Calibrated Submanifolds and Connections'' and the Banff International Research Center for their hospitality.

\bibliographystyle{unstr}

\begin{thebibliography}{99}
\bibitem{Haydys:2010dv}
 A.~Haydys,
 ``Fukaya-Seidel Category and Gauge Theory,''
 arXiv:1010.2353 [math.SG].
  
\bibitem{Witten:2011zz}
  E.~Witten,
  ``Fivebranes and Knots,''
Quantum\ Topology\ {\bf 3} (2012) 1-137 
  [arXiv:1101.3216 [hep-th]].
   
\bibitem{Witten:2011pz} 
  E.~Witten,
  ``Khovanov Homology and Gauge Theory,''
   in R. Kirby, V. Krushkal, and Z. Wang, eds., {\em Proceedings Of The FreedmanFest} (Mathematical Sciences Publishers, 2012), 
  arXiv:1108.3103 [math.GT].

\bibitem{Khovanov99}
M.~Khovanov, 
``A Categorification of the Jones Polynomial,''
Duke\ Math.\ J.\ {\bf 101} (2000), no. 3, 359Ð426 [arXiv:math/9908171]. 

\bibitem{Khovanov01}
M.~Khovanov, 
``A Functor-valued Invariant of Tangles,''
Algebr.\ Geom.\ Topol.\ {\bf 2} (2002), 665Ð741 [arXiv:math/0103190].

  
\bibitem{P}
V.~Ya.~Pidstrigach,  
``Hyper-K\"ahler Manifolds and the Seiberg-Witten Equations,'' Tr.\ Mat.\ Inst.\ Steklova {\bf 246} (2004), Algebr.\ Geom.\ Metody,\ Svyazi\ i\ Prilozh., 263--276; translation in Proc.\ Steklov\ Inst.\ Math.\ 2004, no. 3 {\bf 246}, 249Ð262.

\bibitem{Gukov:2004hz}
  S.~Gukov, A.~S.~Schwarz, and C.~Vafa,
  ``Khovanov-Rozansky Homology and Topological Strings,''
  Lett.\ Math.\ Phys.\  {\bf 74}, 53-74 (2005).
  [hep-th/0412243].  
  
\bibitem{Ooguri:1999bv} 
  H.~Ooguri and C.~Vafa,
  ``Knot invariants and topological strings,''
  Nucl.\ Phys.\ B {\bf 577}, 419 (2000)
  [hep-th/9912123].
  
\bibitem{Kapustin:2006pk} 
  A.~Kapustin and E.~Witten,
  ``Electric-Magnetic Duality and the Geometric Langlands Program,''
  Commun.\ Num.\ Theor.\ Phys.\  {\bf 1}, 1 (2007)
  [hep-th/0604151].
  
\bibitem{Vafa:1994tf} 
  C.~Vafa and E.~Witten,
  ``A Strong Coupling Test of S Duality,''
  Nucl.\ Phys.\ B {\bf 431}, 3 (1994)
  [hep-th/9408074].
  
\bibitem{CDFN}
E.~Corrigan, C.~Devchand, D.~B.~Fairlie, and J.~Nuyts,
``First-order Equations for Gauge Fields in Spaces of Dimension Greater that Four,''
Nucl.\ Phys.\ {\bf B214} (1983) 452-464.

\bibitem{Ward}
R.~S.~Ward, 
``Completely Solvable Gauge-field Equations in Dimension Greater than Four,'' 
Nucl.\ Phys.\ {\bf B236} (1984), 381Ð396.

\bibitem{ReyesCarrion:1998si}
  R.~Reyes Carrion,
  ``A Generalization of the Notion of Instanton,''
  Differ.\ Geom.\ Appl.\  {\bf 8}, 1-20 (1998).  
  
\bibitem{Donaldson:1996kp}
  S.~K.~Donaldson and R.~P.~Thomas,
  ``Gauge Theory in Higher Dimensions,''
in proceedings of 
{\it Conference on Geometric Issues in Foundations of Science in honor of Sir Roger Penrose's 65th Birthday, Oxford, England, 25-29
Jun 1996}, 
Editors: S.A. Huggett, L.J. Mason, K.P. Tod, S.T. Tsou. Oxford Univ. Pr., 1998. 431p.
  \href{http://www.slac.stanford.edu/spires/find/hep/www?irn=4956141}{SPIRES entry}


\bibitem{Donaldson:2009yq}
  S.~Donaldson and E.~Segal,
  ``Gauge Theory in Higher Dimensions, II,''
Geometry of special holonomy and related topics, 1Ð41, Surv.\ Differ.\ Geom.\ 16, Int. Press, Somerville, MA, 2011 
  [arXiv:0902.3239 [math.DG]].


\bibitem{Salamon:2010cs}
  D.~A.~Salamon and T.~Walpuski,
  ``Notes on the Octonians,''
  [arXiv:1005.2820 [math.RA]].

  
\bibitem{Diaconescu:1996rk}
  D.~E.~Diaconescu,
  ``D-branes, Monopoles and Nahm Equations,''
  Nucl.\ Phys.\  B {\bf 503}, 220 (1997)
  [arXiv:hep-th/9608163].
  
\bibitem{Gaiotto:2011nm} 
  D.~Gaiotto and E.~Witten,
  ``Knot Invariants from Four-Dimensional Gauge Theory,''
  Adv.\ Theor.\ Math.\ Phys.\  {\bf 16}, no. 3, 935 (2012)
  [arXiv:1106.4789 [hep-th]].
  
\bibitem{FairlieNuyts84} 
D.~B.~Fairlie and J.~Nuyts,
``Spherically Symmetric Solutions of Gauge Theories in Eight Dimensions,''
J.\ Phys.\ A: Math. Gen. {\bf 17} (1984) 2867-2872.	

\bibitem{FubiniNicolai85}
S.~Fubini and H.~Nicolai,
``The Octonionic Instanton,''
Physics Letters {\bf B155}   (1985)  369-372.

\bibitem{Dunajski:2011sx} 
  M.~Dunajski and M.~Hoegner,
  ``$SU(2)$ Solutions to Self-duality Equations in Eight Dimensions,''
  J.\ Geom.\ Phys.\ {\bf 62} (2012) 1747-1759 
  [arXiv:1109.4537 [hep-th]].

\bibitem{Fairlie:1984bh} 
  D.~B.~Fairlie and J.~Nuyts,
  ``Integration Conditions for First Order Differential Linear Equations in Higher Dimensional Gauge Theories,''
  J.\ Math.\ Phys.\  {\bf 25}, 2025 (1984).

\bibitem{Mazzeo:2013zga} 
  R.~Mazzeo and E.~Witten,
  ``The Nahm Pole Boundary Condition,''
  arXiv:1311.3167 [math.DG].
  
\bibitem{Witten:2014xwa} 
  E.~Witten,
  ``Two Lectures on the Jones Polynomial and Khovanov Homology,''
  arXiv:1401.6996 [math.GT].    
  
\bibitem{Witten:1982im} 
  E.~Witten,
  ``Supersymmetry and Morse Theory,''
  J.\ Diff.\ Geom.\  {\bf 17}, 661 (1982).
    
\bibitem{Acharya:1997gp} 
  B.~S.~Acharya, M.~O'Loughlin, and B.~J.~Spence,
  ``Higher Dimensional Analogs of Donaldson-Witten Theory,''
  Nucl.\ Phys.\ B {\bf 503}, 657 (1997)
  [hep-th/9705138].
  
\bibitem{Witten:1988ze} 
  E.~Witten,
  ``Topological Quantum Field Theory,''
  Commun.\ Math.\ Phys.\  {\bf 117}, 353 (1988).
  



\end{thebibliography}

\end{document}